\pdfoutput=1
\RequirePackage{ifpdf}
\ifpdf 
\documentclass[pdftex]{sigma}
\else
\documentclass{sigma}
\fi

\DeclareMathOperator{\vol}{vol}

\begin{document}

\allowdisplaybreaks

\renewcommand{\thefootnote}{$\star$}

\renewcommand{\PaperNumber}{037}

\FirstPageHeading

\ShortArticleName{Twistor Theory of the Airy Equation}

\ArticleName{Twistor Theory of the Airy Equation\footnote{This paper is a~contribution to the Special Issue on Progress
in Twistor Theory. The full collection is available at
\href{http://www.emis.de/journals/SIGMA/twistors.html}{http://www.emis.de/journals/SIGMA/twistors.html}}}

\Author{Michael COLE and Maciej DUNAJSKI}

\AuthorNameForHeading{M.~Cole and M.~Dunajski}

\Address{Department of Applied Mathematics and Theoretical Physics, University of Cambridge,\\
Wilberforce Road, Cambridge CB3 0WA, UK}

\Email{\href{mailto:m.dunajski@damtp.cam.ac.uk}{m.dunajski@damtp.cam.ac.uk}}

\ArticleDates{Received November 28, 2013, in f\/inal form March 18, 2014; Published online March 29, 2014}

\Abstract{We demonstrate how the complex integral formula for the Airy functions arises from Penrose's twistor contour
integral formula.
We then use the Lax formulation of the isomonodromy problem with one irregular singularity of order four to show that
the Airy equation arises from the anti-self-duality equations for conformal structures of neutral signature invariant
under the isometric action of the Bianchi~II group.
This conformal structure admits a~null-K\"ahler metric in its conformal class which we construct explicitly.}

\Keywords{twistor theory; Airy equation; self-duality}

\Classification{32L25; 34M56}

\renewcommand{\thefootnote}{\arabic{footnote}}
\setcounter{footnote}{0}

\section{Introduction}

\begin{figure}[h!]
\centering \includegraphics[width=5cm]{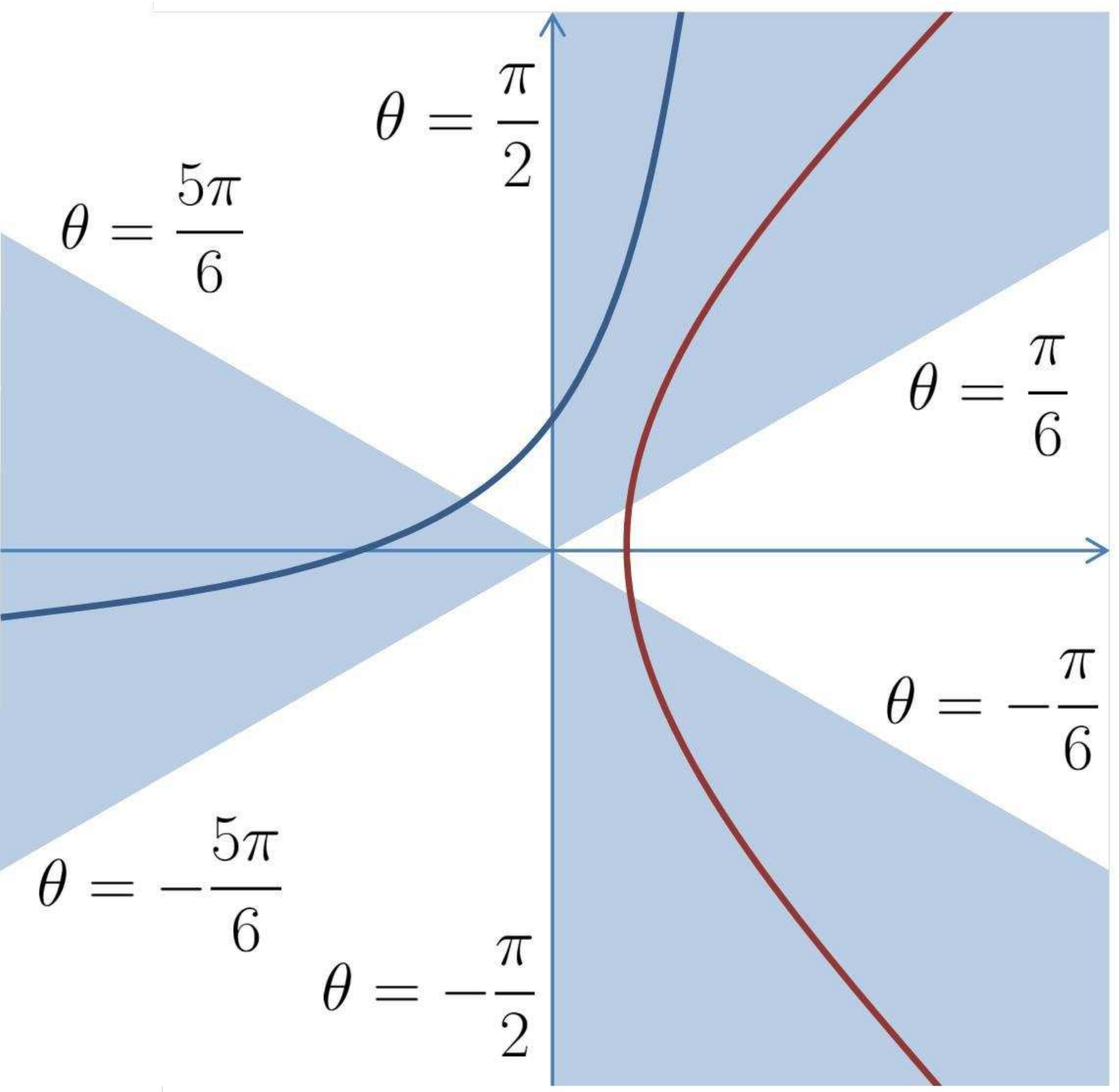}

\caption{Contours in Airy's formula start and end in the shaded sectors.}\label{Fig1}
\end{figure}

The Airy equation
\begin{gather}
\label{airy}
f(t)''+tf(t)=0
\end{gather}
admits solutions given by integrals
\begin{gather}
\label{airy_1}
f(t)=\int_\Gamma \exp{\left(\frac{1}{3}\lambda^3+\lambda t \right)}d\lambda,
\end{gather}
where $\lambda\in{\mathbb{C}}$ is an auxiliary complex parameter (see, e.g.,~\cite{Airy}).
Two linearly independent solutions correspond to open contours $\Gamma=\Gamma_1$ and $\Gamma=\Gamma_2$ in the complex
plane, each of which begins in one of the three shaded sectors where $\cos{(3\operatorname{arg}(\lambda))}<0$ and ends in
another such sector (Fig.~\ref{Fig1}).

This observation can be verif\/ied by dif\/ferentiating inside the integral and integrating by parts.
The boundary term is then required to vanish which restricts the allowed form of the contours.

Airy's formula~\eqref{airy_1} is reminiscent of Penrose's twistor contour integral formula~\cite{penrose} for the
solutions to the wave equation on the complexif\/ied Minkowski space ${\mathbb{CM}}={\mathbb{C}}^4$.
If $(w, z, \tilde{w}, \tilde{z})$ are holomorphic coordinates on $\mathbb{CM}$ such that the line element is
${ds}^2=dzd\tilde{z}-dwd\tilde{w}$, then all solutions of the wave equation
\begin{gather}
\label{wave}
\phi_{z\tilde{z}}-\phi_{w\tilde{w}}=0
\end{gather}
are of the form
\begin{gather}
\label{penrose_1}
\phi(w, z, \tilde{w}, \tilde{z})=\oint_\Gamma \psi(w+\lambda \tilde{z}, z+\lambda \tilde{w}, \lambda)d\lambda,
\end{gather}
where $\Gamma\subset{\mathbb{CP}}^1$ is a~closed contour, and the function $\psi$ is holomorphic in
$\lambda\in{\mathbb{CP}}^1$ except some singularities inside~$\Gamma$.

In Section~\ref{section1} we shall demonstrate that~\eqref{airy_1} is a~special case of~\eqref{penrose_1}.
Our procedure is an application of symmetry reductions~\cite{mason} which have previously been used to analyse
non-linear ODEs arising from a~non-abelian generalisation of~\eqref{wave}.
Our result emphasises the cohomological nature of twistor integral formulae: the general solution to~\eqref{wave}
depends on two arbitrary functions of three variables, and yet the twistor formula~\eqref{penrose_1} contains only one
such function on the r.h.s.
How is that possible? The answer~-- hidden in the abstract machinery of the Penrose transform~\cite{eastwood}~-- is that
solutions to~\eqref{wave} correspond to the equivalence classes $(\Gamma, \psi)$ of contours and cohomology
representatives rather than (naively) to holomorphic functions of three variables.
The example of the Airy equation makes this explicit: the general solution to the linear ODE~\eqref{airy} depends on two
constants of integration, and is a~linear combination of two linearly independent solutions.
Both solutions are of the form~\eqref{airy_1} with the same integrand (which plays a~role of the twistor function~$\psi$), but dif\/ferent contours.

In Section~\ref{section2} we shall show that the Airy equation also arises from the anti-self-duality equations on
a~cohomogeneity one-conformal structure of neutral signature invariant under the Bianchi~II group.
This conformal structure admits a~null-K\"ahler metric in its conformal class which we construct explicitly.
The twistor distribution def\/ining the $\alpha$-surfaces is equivalent to the Lax pair of~\cite{Jimbo} for the
isomonodromic problem with one irregular singularity of order four.

\section{Construction of the twistor function}\label{section1}

Let us start from the following lemma which can be proven by explicit calculation:
\begin{lemma}
\label{lemma1}
Set $t=\tilde{z}-z-\tilde{w}^2$.
The function
\begin{gather}
\label{group_invariant}
\phi(w, z, \tilde{w}, \tilde{z})=\exp \left(w-\tilde{w}(z-\tilde{z})-\frac{2}{3}\tilde{w}^3\right) f(t)
\end{gather}
satisfies the wave equation~\eqref{wave} if and only if $f(t)$ is a~solution to the Airy equation~\eqref{airy}.
\end{lemma}

This result is valid both in the holomorphic and real category.
In the latter case all the coordinates are assumed to be real, and~$\phi$ satisf\/ies the ultra-hyperbolic wave equation
on ${\mathbb{R}}^{2, 2}$~\cite{john, woodhouse}.
The geometric interpretation of Lemma~\ref{lemma1} is as follows: The wave equation on $\mathbb{CM}$ is conformally
invariant, and the conformal group of the complexif\/ied Minkowski space is ${\rm SL}(4, {\mathbb{C}})$.
The special solutions~\eqref{group_invariant} are invariant under the action of the three-dimensional abelian subgroup~$H$ of ${\rm SL}(4, {\mathbb{C}})$ generated by the vector f\/ields
\begin{gather}
\label{vector_fields}
X=\partial_w,
\qquad
Y=\partial_z+\partial_{\tilde{z}},
\qquad
Z=(z-\tilde{z})\partial_w+\tilde{w}(\partial_{\tilde{z}} -\partial_z)+\partial_{\tilde{w}},
\end{gather}
and invariance condition takes the form
\begin{gather*}
X(\phi)=\phi,
\qquad
Y(\phi)=0,
\qquad
Z(\phi)=0.
\end{gather*}
The readers familiar with the twistor approach to the Painlev\'e equations~\cite{mason} will recognise this abelian
group.
A symmetry reduction of the anti-self-dual Yang--Mills equation by $H$ yields the Painlev\'e~II equation.
This is not unexpected, as from the isomonodromic perspective both the Airy equation and the Painlev\'e~II correspond to
an irregular singular point of order four.

To proceed further we recall the basic twistor correspondence and the double f\/ibration picture (see, e.g.,~\cite{Dunajski_book,mason}).
The points of the three-dimensional twistor space ${\mathbb{PT}}={\mathbb{CP}}^3-{\mathbb{CP}}^1$ are $\alpha$-planes in
$\mathbb{CM}$: two-dimensional planes which are totally null with respect to the line element~$ds^2$, and such that
their tangent bi-vector is self-dual with respect to a~holomorphic volume form $dw\wedge d\tilde{w}\wedge dz\wedge
d\tilde{z}$.
There is a~rational curve $L_p\cong{\mathbb{CP}}^1$ worth of such planes through each point $p\in{\mathbb{CM}}$, and so
points in $\mathbb{CM}$ correspond to rational curves in $\mathbb{PT}$.
The twistor space arises as a~quotient of the f\/ive-dimensional correspondence space
$\mathcal{F}={\mathbb{CM}}\times{\mathbb{CP}}^1$ by a~two-dimensional integrable distribution spanned by the vector f\/ields
\begin{gather*}
l=\lambda_0\partial_{\tilde{z}}-\lambda_1\partial_w,
\qquad
m=\lambda_0\partial_{\tilde{w}}-\lambda_1\partial_z,
\end{gather*}
where $[\lambda_0, \lambda_1]$ are homogeneous coordinates of a~point in ${\mathbb{CP}}^1$.
For each f\/ixed $[\lambda_0, \lambda_1]\in {\mathbb{CP}}^1$ these vector f\/ields span an $\alpha$-plane through a~point in
$\mathbb{CM}$.
A twistor function is a~function on~$\mathcal{F}$ which is constant along~$l$ and~$m$.
In a~patch containing $[1,0]\in{\mathbb{CP}}^1$ def\/ine $\lambda=\lambda_1/\lambda_0$.
The local coordinates on~$\mathbb{PT}$ pull back to three twistor functions
\begin{gather}
\label{line}
\mu=w+\lambda \tilde{z},
\qquad
\nu=z+\lambda \tilde{w},
\qquad
\lambda.
\end{gather}
Fixing $(w, z, \tilde{w}, \tilde{z})$ in~\eqref{line} gives a~rational curve in $\mathcal{F}$ which descends down to
$\mathbb{PT}$.
Conversely, f\/ixing $(\lambda, \mu, \nu)$ in~\eqref{line} gives an $\alpha$-plane in $\mathbb{CM}$.

To arrive at the contour formula~\eqref{penrose_1} consider a~twistor function $\psi(\mu, \nu, \lambda)$ def\/ined on an
intersection of two open sets containing $\lambda_0=0$ and $\lambda_1=0$ respectively.
Now restrict $\psi$ to a~twistor line $L_p\cong {\mathbb{CP}}^1$, and integrate its pull back to $\mathcal{F}$ over
a~closed contour in the pre-image of $L_p$ in $\mathcal{F}$.
This gives a~function at $p\in {\mathbb{CM}}$ which satisf\/ies the wave equation~\eqref{wave}.
The volume element $d\lambda$ on $L_p$ is a~section of a~line bundle $\mathcal{O}(2)\rightarrow L_p$, so for all of this
to work $\psi$ must really be a~representative of a~cohomology class $H^1(L_p, \mathcal{O}(-2))$.

To implement the symmetry condition we need to consider the action of $H$ on $\mathcal{F}$, where in addition to moving
points in $\mathbb{CM}$ it also changes the $\alpha$-planes through a~point.
This is done by constructing lifts $(X'', Y'', Z'')$ of the generators~\eqref{vector_fields} to $\mathcal{F}$ such that
the resulting vector f\/ields commute with $l$ and $m$ modulo $(l, m)$.
The lifted vector f\/ields are
\begin{gather*}
X''=X,
\qquad
Y''=Y,
\qquad
Z''=Z+\partial_\lambda.
\end{gather*}
They push forward to holomorphic vector f\/ields $(X', Y', Z')$ on the twistor space, which can be constructed by an
application of the chain rule.
This yields
\begin{gather*}
X'=\partial_\mu,
\qquad
Y'=\partial_\nu+\lambda\partial_\mu,
\qquad
Z'=\nu\partial_\mu+\lambda\partial_\nu+\partial_\lambda.
\end{gather*}
Now set $\Omega=\psi(\mu, \nu, \lambda)d\lambda$, and impose the symmetry condition
\begin{gather*}
{\mathcal L}_{X'}\Omega=\Omega,
\qquad
{\mathcal L}_{Y'}\Omega=0,
\qquad
{\mathcal L}_{Z'}\Omega=0,
\end{gather*}
where ${\mathcal L}$ is the Lie derivative.
This yields
\begin{gather*}
\Omega=\exp \big(\mu-\lambda\nu+\lambda^3/3\big) d\lambda
=\exp \left(w-\tilde{w}(z-\tilde{z})-\frac{2}{3}\tilde{w}^3\right)\! \exp{\left(\frac{1}{3}(\lambda-\tilde{w})^3+(\lambda-\tilde{w})
t\right)}d\lambda
\end{gather*}
up to an overall constant multiple, where in the second line above we have restricted $\Omega$ to the twistor
line~\eqref{line}.
Performing a~M\"obius transformation $\lambda\rightarrow (\lambda+\tilde{w})$ and using~\eqref{group_invariant} reduces
the Penrose formula~\eqref{penrose_1} to the Airy formula~\eqref{airy_1}.
The contours from Fig.~\ref{Fig1} are now re-interpreted as closed contours in the Riemann sphere $L_p\subset
\mathbb{PT}$ corresponding to $p\in\mathbb{CM}$.
The twistor function~$\psi$ has an essential singularity at the pole $\lambda=\infty$ which belongs to both contours.
The integrals are nevertheless well def\/ined as~$\psi$ approaches the singularity in the shaded sectors.
\begin{figure}[t]
\centering
\includegraphics[width=4.5cm]{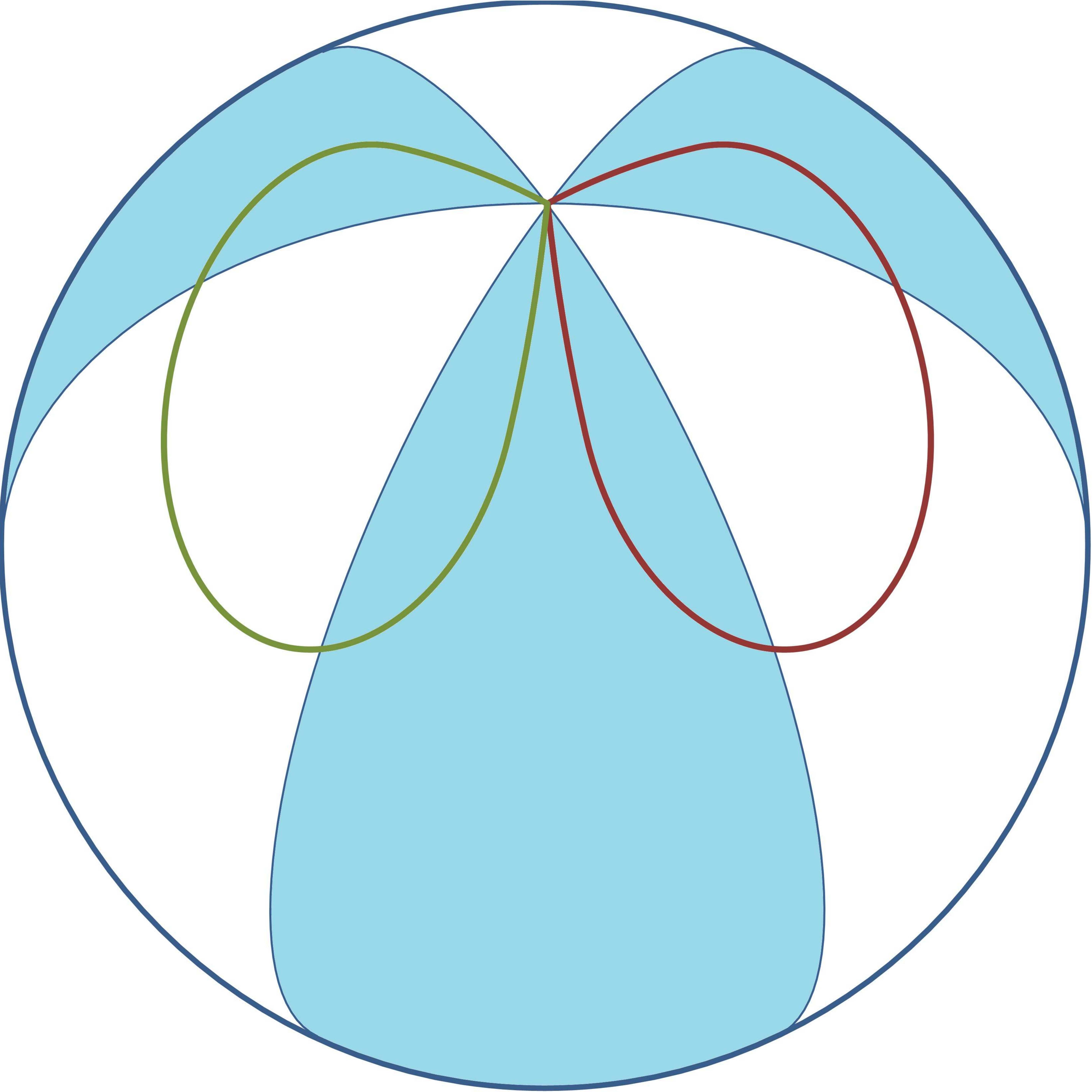}
\caption{Closed contours in the Riemann sphere corresponding to a~point in $\mathbb{CM}$.\label{Fig2}}
\end{figure}

\section{ASD conformal structure with Bianchi~II symmetry}\label{section2}

In this section
we shall give an alternative twistor construction for the Airy equation, this time using this equation
to construct a~conformal class of curved metrics in four dimensions with anti-self-dual Weyl tensor.

The nonlinear graviton construction on Penrose~\cite{penrose_NG} extends twistor theory to curved backgrounds.
An $\alpha$-surface in an oriented holomorphic four-manifold $M$ with a~holomorphic metric $g$ is a~two-dimensional
surface $\xi$ such that $T_p\xi$ is an $\alpha$-plane for any $p\in\xi$.
A seminal result of~\cite{penrose_NG} is that there locally exist a~three-parameter family of $\alpha$-surfaces if\/f
the self-dual part of the Weyl tensor of $g$ vanishes.
We then say that the conformal structure $[g]:=\{\Omega g, \Omega:M\rightarrow {\mathbb{R}}^+\}$ is anti-self-dual
(ASD).
A twistor space $\mathbb{PT}$ of an ASD four-manifold is def\/ined to be a~space of $\alpha$-surfaces.
It is a~three-dimensional complex manifold with a~four-parameter family of rational curves with normal bundle
$\mathcal{O}(1)\oplus\mathcal{O}(1)$.
The conformal structure of $g$ is encoded in the complex structure of the twistor space, which for non-conformally f\/lat
ASD manifolds is dif\/ferent than that of ${\mathbb{CP}}^3-{\mathbb{CP}}^1$.
A convenient way to express the ASD condition on a~conformal structure is summarised in the following
theorem~\cite{Dunajski_book,mason}.
\begin{theorem}
\label{laxprop}
Let $Z$, $W$, $\widetilde{Z}$, $\widetilde{W}$ be four independent holomorphic vector fields on~$M$.
The conformal structure defined by
\begin{gather}
\label{tetrad_0}
g=Z\odot \widetilde{Z}- W\odot \widetilde{W}
\end{gather}
is ASD if and only if there exists functions $f_0$, $f_1$ on $M \times {\mathbb{CP}}^1$ cubic in $\lambda\in {\mathbb{CP}}^1$
such that the distribution{\samepage
\begin{gather}
\label{tetradlax}
l = \widetilde{Z} - \lambda W + f_0 \frac{\partial}{\partial \lambda},
\qquad
m = \widetilde{W} - \lambda Z + f_1 \frac{\partial}{\partial \lambda}
\end{gather}
is Frobenius integrable, i.e.~$[l,m] = 0$ modulo $l$ and $m$.}
\end{theorem}

If the integrability condition holds then there is a~${\mathbb{CP}}^1$-worth of $\alpha$-surfaces spanned by $\{
\widetilde{Z} - \lambda W, \widetilde{W} - \lambda Z\}$ through any point in~$M$.
If all vectors $(Z, \dots, \widetilde{W})$ are real then the signature of~$g$ is~$(2, 2)$, and there exists an
${\mathbb{RP}}^1$-worth of real $\alpha$-surfaces through each point.

From now on we shall additionally assume that there exists a~three-dimensional Lie group~$G$ acting on~$M$ by conformal
isometries with generically three-dimensional orbits.
The conformal isometries are generated by the right-invariant vector f\/ields $R_j$, $j=1, 2, 3$ on~$G$.
The $G$-action on $M$ maps $\alpha$-surfaces to $\alpha$-surfaces and thus gives rise to a~group action of~$G$ on the
twistor space~${\mathbb{PT}}$.
We can choose to work in an invariant frame, where the lifts of~$R_j$s to the correspondence space are given by~$R_j$s.
Consider a~quartic
\begin{gather}
\label{quartic}
s = (d \lambda \wedge \vol) (l, m, {R}_1, {R}_2, {R}_3),
\end{gather}
where $\vol$ is the holomorphic volume form on~$M$ such that $\vol( W, \widetilde{W}, Z, \widetilde{Z})=1$.
To make contact with the isomonodromic problem for the Airy equation we shall assume that~$s$ vanishes at one value of~$\lambda$ to order four.
We can write the null tetrad of Theorem~\ref{laxprop} terms of the vector f\/ield~$\partial_t$ orthogonal to the
$G$-orbits, and three linearly independent vector f\/ields~$P$, $Q$, $R$ tangent to~$G$ which are $t$-dependent and
invariant under left translations:
\begin{gather}
\label{tetrad}
\widetilde{Z} = \partial_t + {R},
\qquad
Z = \partial_t - {R},
\qquad
W = -P,
\qquad
\widetilde{W}=Q.
\end{gather}
Moreover the invariance condition implies that $f_0$ and $f_1$ are constant on $G$, and so depend only on $\lambda$ and
$t$.
A direct calculation now shows that the quartic $s$ is proportional to $(\lambda f_0 + f_1)$.
This quartic has a~quadrupole zero which we shall move to $\lambda=\infty$ by a~M\"obius transformation.
Using the freedom in the left and right rotations of the tetrad, and the ASD equations it is possible to set $(f_0=0$,
$f_1=1)$.
Now consider\footnote{In~\cite{mmw} it was instead assumed that the quartic $s$ has four distinct zeros, and that
$G={\rm SL}(2, {\mathbb{C}})$ which lead to the isomonodromic Lax pair~\cite{Jimbo} for Painlev\'e~VI.} a~pair of linear
combinations of $l$ and $m$~\eqref{tetradlax} given by
\begin{gather*}
L:= \frac{\lambda l + m}{\lambda f_0 + f_1} = \frac{\partial}{\partial \lambda} + Q+2\lambda R+\lambda^2 P,
\qquad
M:= \frac{f_1 l - f_0 m}{\lambda f_0 + f_1} = \frac{\partial}{\partial t} + R+\lambda P.
\end{gather*}
Since the conformal class is ASD, Theorem~\ref{laxprop} implies that $[L, M] =0$, modulo $L$ and $M$.
However $[L,M]$ does not contain $\partial_\lambda$ or $\partial_t$, thus $[L,M]$ must be identically zero which yields
\begin{gather}
\label{isom_airy}
Q'=[Q, R]+P,
\qquad
R'=\frac{1}{2}[Q, P],
\qquad
P'=[R, P].
\end{gather}
We shall now make a~choice for $G$, and take it to be the Bianchi~II group.
Its Lie algebra is generated by the left invariant vector f\/ields $L_j$ on $G$ which satisfy
\begin{gather}
\label{bianchi_2}
[L_1, L_2]=L_1,
\qquad
[L_1, L_3]=0,
\qquad
[L_2, L_3]=0.
\end{gather}
The connection with the Airy equation is provided by the following result.
\begin{theorem}
\label{lemma2}
A general cohomogeneity-one Bianchi~II ASD conformal class such that the twistor quartic~\eqref{quartic} admits a~zero
of order four is of the form~\eqref{tetrad_0},~\eqref{tetrad}, with
\begin{gather}
\label{ansatz}
P=\frac{1}{2}FL_1+L_2,
\nonumber
\\
Q=\left(\frac{1}{2}({c_1}^2+t)F+c_1F'+F''\right)L_1+({c_1}^2+t)L_2+c_2 L_3,
\\
R=\frac{1}{2}(c_1F+F')L_1+c_1L_2+L_3,
\nonumber
\end{gather}
where $(c_1, c_2)$ are constants, and $f(t)=F'(t)$ satisfies the Airy equation~\eqref{airy}.
\end{theorem}
\begin{proof}
To complete the proof we need to show that any solution to the reduced ASD equations~\eqref{isom_airy} can be put in the
form~\eqref{ansatz} without the loss of generality.
The three vector f\/ields~$P$, $Q$, $R$ can be written in the basis of left-invariant vector f\/ields $L_1$, $L_2$, $L_3$
satisfying~\eqref{bianchi_2} as
\begin{gather*}
P=p_1(t) L_1 + p_2(t) L_2 + p_3(t) L_3,
\\
Q=q_1(t) L_1 + q_2(t) L_2 + q_3(t) L_3,
\\
R=r_1(t) L_1 + r_2(t) L_2 + r_3(t) L_3,
\end{gather*}
for some functions $p_j(t)$, $q_j(t)$, $r_j(t)$.
The Lie algebra relation~\eqref{bianchi_2} is preserved by
\begin{gather}
\label{scale_L}
L_1\rightarrow \alpha L_1,
\qquad
L_2\rightarrow L_2+\beta L_3,
\qquad
L_3\rightarrow \gamma L_3,
\end{gather}
where ($\alpha\neq 0$, $\beta$, $\gamma\neq 0$) are constants.
The third equation in~\eqref{isom_airy} implies that $p_2=c_0$ is a~constant and that $p_3$ is a~constant which can be
set to zero using~\eqref{scale_L}.
We also get $p_1'=r_1p_2-p_1r_2$.
Let us set $p_1=F/2$, where $F=F(t)$.
The second equation in~\eqref{isom_airy} implies that $r_2=c_1$ is a~constant and that $r_3$ is another constant which
we can set to $1$ using~\eqref{scale_L}.
We also obtain
\begin{gather*}
q_1=\frac{2{r_1}'}{c_0}+\frac{q_2 F}{2 c_0}.
\end{gather*}
Finally the f\/irst equation in~\eqref{isom_airy} implies that $q_3=c_2$ is a~constant, ${q_2}=c_0t+c_3$, where $c_3$ is
another constant, and it also gives
\begin{gather*}
F'''+\big(\big(c_3c_0-{c_1}^2\big)+t{c_0}^2\big)F'=0.
\end{gather*}
The result now follows by making an af\/f\/ine transformation of $t$ to set $c_0=1$, $c_3={c_1}^2$.
\end{proof}

We shall end this section
discussing the connection between the system~\eqref{isom_airy} and the isomo\-nodromic problem
with irregular singularity of order four.
Consider a~$2\times 2$ matrix
\begin{gather*}
\Theta(t, \lambda)=Q+2\lambda R+\lambda^2 P,
\end{gather*}
where $\lambda\in{\mathbb{CP}}^1$, and $P$, $Q$, $R$ are elements of a~matrix Lie algebra~$\mathfrak{g}$ which also depend on
a~parameter $t$.
For a~chosen f\/ixed value of $t$ consider a~linear matrix ODE
\begin{gather*}
\frac{d \Psi}{d \lambda}+\Theta \Psi=0.
\end{gather*}
Now allow $t$ to vary on the complex plane, so that the matrix fundamental solution $\Psi$ depends on $\lambda$ and $t$.
The monodromy around the fourth-order pole $\lambda=\infty$ does not depend on $t$ if $\Psi$ satisf\/ies~\cite{Jimbo}
\begin{gather*}
\frac{\partial \Psi}{\partial \lambda}+\Theta\Psi=0,
\qquad
\frac{\partial \Psi}{\partial t}+\Theta_+\Psi=0,
\qquad
\text{where}
\quad
\Theta_+:=R+\lambda P.
\end{gather*}
The compatibility conditions for this overdetermined linear system reduce to system of nonlinear matrix ODEs\footnote{
These compatibility conditions are
\begin{gather*}
\frac{1}{2}\frac{\partial^2 \Theta}{\partial \lambda^2}-\frac{\partial \Theta}{\partial t}
+\frac{1}{2}\left[\Theta,\frac{\partial \Theta}{\partial \lambda}\right]=0.
\end{gather*}
Rescaling $\partial/\partial t \rightarrow \epsilon \partial/\partial t$ and $\partial/\partial \lambda\rightarrow
\epsilon\partial/\partial \lambda$, and taking the dispersionless limit $\epsilon\rightarrow 0$ yields the Nahm
equations for $(P, Q, R)$.
In this limit the spectral curve $ S=\{(\omega, \lambda)\in{ T{\mathbb{CP}}^1}\mid \det({\bf 1}\omega-\Theta(t,\lambda))=0\}$
does not depend on $t$.
This is in agreement with the observation of~\cite{takasaki} that isospectral deformations arise as a~limit of
isomonodromic deformations.}~\eqref{isom_airy} for $(P, Q, R)$.
If $P$ is diagonalisable, and $\mathfrak{g}=\mathfrak{sl}(2, {\mathbb{C}})$, then~\eqref{isom_airy} reduce to Painlev\'e~II.
Theorem~\ref{lemma2} shows that if instead $\mathfrak{g}$ is the Bianchi~II algebra~\eqref{bianchi_2} then the
isomonodromic condition is the (derivative of) the Airy equation.

\subsection{ASD null-K\"ahler structure}

A null-K\"ahler structure on a~four-dimensional manifold $M$ is a~pair
$(\hat{g}, N)$ where $\hat{g}$ is a~metric of signature $(2, 2)$ and $N: TM\longrightarrow TM$ is a~rank-$2$
endomorphism such that
\begin{gather*}
N^2=0,
\qquad
\hat{g}(NX, Y)+\hat{g}(X, NY)=0,
\qquad
\nabla N=0
\end{gather*}
for all vector f\/ields $X$, $Y$ on $M$.
Given such $N$ and $\hat{g}$ we can construct a~null-K\"ahler two-form~$\Sigma$ such that $\Sigma(X, Y)=\hat{g}(NX, Y)$
and
\begin{gather*}
\nabla\Sigma=0,
\qquad
\Sigma\wedge\Sigma=0.
\end{gather*}
In~\cite{D02} it was shown that a~null-K\"ahler structure gives rise to a~preferred section of $\kappa^{-1/4}$, where~$\kappa$ is the holomorphic canonical bundle of the twistor space, and conversely given that~$\mathbb{PT}$ admits such
an anti-canonical divisor the corresponding ASD conformal class admits a~null-K\"ahler structure.
The conformal class def\/ined by $(P, Q, R)$ from Lemma~\ref{lemma2} gives rise to a~section of~$\kappa^{-1/4}$ given by
a~push-forward of the quartic~\eqref{quartic} to $\mathbb{PT}$ vanishes at each twistor line at one point, where the
holomorphic vector f\/ields corresponding to the isometries become linearly dependent.
This conformal class should therefore contain a~null-K\"ahler structure.
To construct it explicitly, choose the coordinates $(x, y, z)$ on~$G$ such that the left-invariant vector f\/ields are
\begin{gather*}
L_1=e^{-y}{\partial_x},
\qquad
L_2={\partial_y},
\qquad
L_3={\partial_z}.
\end{gather*}
To simplify the formulae, we shall also choose $c_1=c_2=0$ in~\eqref{ansatz}.
This yields the basis of one-forms dual to $(Z, W, \widetilde{Z}, \widetilde{W})$ given by
\begin{gather*}
{\bf e}_{\widetilde{Z}}=\frac{1}{2}(dt+dz),
\qquad
{\bf e}_{\widetilde{W}}= \frac{1}{F''}\left(e^ydx-\frac{1}{2}Fdy-\frac{1}{2}F'dz\right),
\\
{\bf e}_{{Z}}=\frac{1}{2}(dt-dz),
\qquad
{\bf e}_W=\frac{1}{F''}\left(te^ydx-\left(\frac{t}{2}F+F''\right)dy- \frac{t}{2}F' dz\right).
\end{gather*}
The conformally rescaled metric
\begin{gather*}
\hat{g}=\Omega\big({\bf e}_{{Z}} \odot {\bf e}_{\widetilde{Z}}- {\bf e}_{{W}}\odot {\bf e}_{\widetilde{W}}\big),
\qquad
\text{where}
\qquad
\Omega=e^{-y}(F'')
\end{gather*}
is ASD and null-K\"ahler with the null-K\"ahler two-form $\Sigma$ given by
\begin{gather*}
\Sigma=\Omega\big({\bf e}_{\widetilde{Z}}\wedge {\bf e}_{\widetilde{W}}\big).
\end{gather*}
This is also gives the null-K\"ahler two-form in the general case when the constants $(c_1, c_2)$ are arbitrary, in
which case the conformal factor needs to be replaced by $\Omega=e^{-y}(F''+(c_1-c_2/2)F')$.

\section{Conclusions}

We have shown that the integral formula for the Airy equation is a~special case of Penrose's
twistor integral formula.
Other special functions also admit a~twistor description.
In particular the hyper-geometric function together with its generalisations have been investigated in~\cite{shah}.
All twistor integral formulae for special functions could presumably be obtained from the results of this work by
conf\/luence of singularities of the twistor quartic $\vol_{\mathbb{PT}}(X', Y', Z')$, where~$X'$,~$Y'$,~$Z'$ are
holomorphic vector f\/ields on~$\mathbb{PT}$ which correspond to the generators of the Painlev\'e~VI abelian subgroup of
${\rm SL}(4, {\mathbb{C}})$.
The details of this have not been worked out, but it may be interesting to do so.

The one-form
\begin{gather*}
A=\phi_z d\tilde{w}+\phi_w d\tilde{z}
\end{gather*}
is a~solution of the ASD Maxwell equations $F={-}{*}F$, where $F=dA$, if and only if $\phi$ sa\-tis\-f\/ies~\eqref{wave}.
ASD Maxwell f\/ields correspond, via the Ward transform~\cite{ward}, to holomorphic line bundles over~$\mathbb{PT}$ which
are trivial on all twistor lines.
Thus there is a~particular two-parameter class of such line bundles corresponding to the solutions of the Airy
equations.
These bundles can be characterised by their $H$-invariance along the lines explained in~\cite{mason}.

In Section~\ref{section2} we have found a~neutral signature cohomogeneity-one metric on a~four-manifold
$M={\mathbb{R}}\times G$, where~$G$ is the three-dimensional Bianchi~II Lie group, such that the anti-self-duality
condition on the Weyl tensor reduces to the Airy equation.
In this case the twistor distribution~$(l, m)$ def\/ining the three-parameter family of $\alpha$-surfaces in $M$ is
equivalent to the Lax pair for the isomonodromic problem with one irregular singularity of order four.
The resulting metric is conformally related to an ASD null-K\"ahler structure.

\subsubsection*{Acknowledgement}

MC would like to thank James Bridgwater for the f\/inancial support.

\pdfbookmark[1]{References}{ref}
\LastPageEnding

\end{document}